\documentclass[fleqn,12pt,twoside]{article}
\usepackage{espcrc1}

\usepackage{graphicx}

\usepackage[figuresright]{rotating}

\newcommand{\AmS}{{\protect\the\textfont2
   A\kern-.1667em\lower.5ex\hbox{M}\kern-.125emS}}

\title{Constraining the Cosmic Star Formation Rate with the MeV Background}

\author{K.Watanabe\address{NASA/GSFC,Code 660.1,Greenbelt, MD 20771, USA}
         D.H.Hartmann\address[CU]{Department of Physics and Astronomy \\
         Clemson University, Clemson, SC 29634-0978, USA},
         M.D.Leising\addressmark[CU] and L.-S.The\addressmark[CU]}

\begin{document}

\maketitle
\begin{abstract}
The Cosmic Gamma-ray Background (CGB) in the MeV regime has been measured 
with COMPTEL
\cite{Wei99} and SMM \cite{Wata97}. The origin of the CGB in this 
energy regime is
believed to be dominated by gamma-rays from Type Ia supernovae. We 
calculate the CGB
spectrum within the framework of FRW cosmology as a function of the 
cosmic star
formation rate, SFR(z) \cite{Wata99}. Several estimates of the SFR(z) 
have been reported 
since the pioneering work of Madau et al. \cite{Mada96}. Here we discuss 
observational
constraints on SFR(z) derived from models of the CGB. In particular, 
we consider the SFR obtained 
from Gamma-Ray Burst observations \cite{Scha02}, which 
increases dramatically
with redshift beyond z $\sim$ 1 in contrast to most estimates which 
saturate or show a mild increase with redshift. Gamma-ray
bursts may be the most powerful tracers of star formation in the 
early universe and thus
provide signposts of the initial epoch of element synthesis. The 
star formation rate implied by GRB statistics results in a gamma-ray 
background that matches the observations more closely than that 
inferred from other tracers of star formation. This may provide some
support for the GRB/SFR-paradigm, which in turn promises a powerful 
diagnostic of star formation, and thus cosmic chemical
evolution, from the era of Population III stars to the present. 
\end{abstract}

\section{The MeV Background}

\begin{figure}[htb]
\includegraphics[angle=90, scale=0.6]{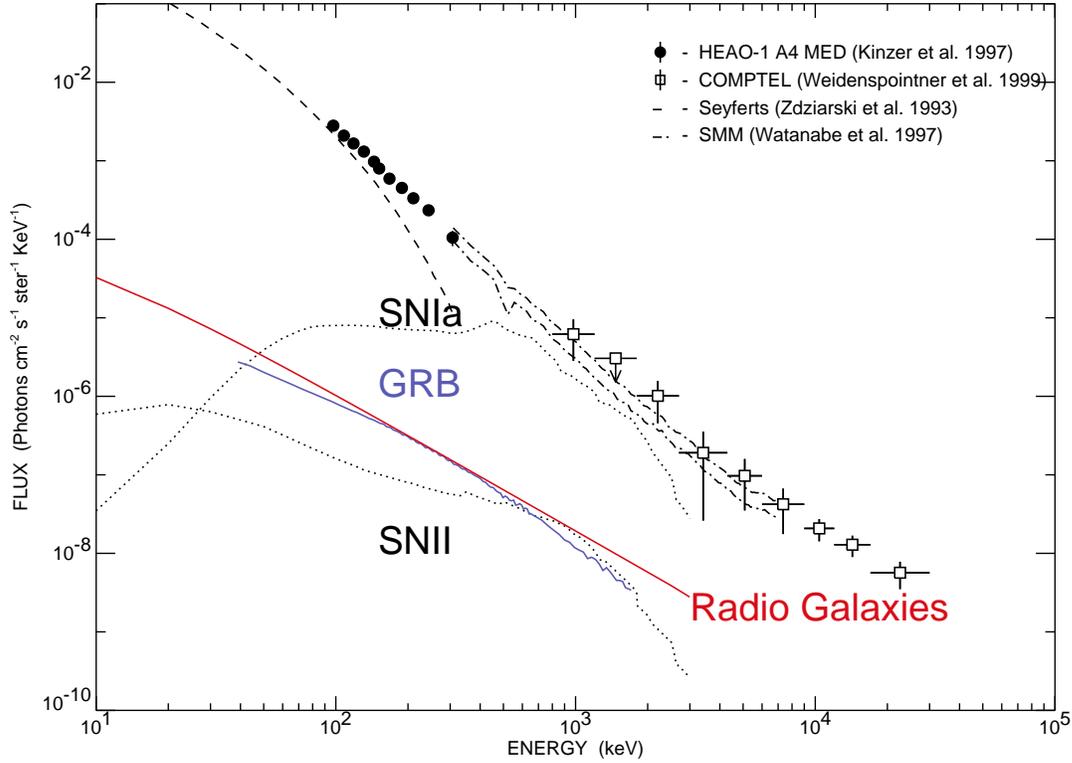}
\caption{Observed MeV-CGB (HEAO-1 A4, SMM, COMPTEL).
SNIa are the dominant contributors in this energy regime,
while SNII, GRBs, and Radio galaxies contribute only a few $\%$ of 
the CGB flux (e.g., \cite{Wata01}).
}
\label{fig:cgb}
\end{figure}

The CGB in the MeV regime has been studied by HEAO-1 A4 \cite{Kinz97},SMM
\cite{Wata97}, and COMPTEL \cite{Wei99}. Supernovae (mostly of type 
Ia) are the dominant
contributors in this energy regime \cite{Wata99}. We previously 
demonstrated that SNII
\cite{Wata99},GRBs, and Cen A-like (FRI) radio galaxies (without 
density evolution
\cite{Wata01}) can account for a few percent of the observed flux. Below 
$\sim$ 300 keV  the
contribution from Seyfert galaxies dominates (e.g.,\cite{Zdzi93}), 
while blazars dominate
above $\sim$ 100 MeV \cite{Sree98}, but see Scharf $\&$ Mukherjee \cite{SM02}
for a discussion of alternatives. Based on the known gamma-ray 
spectra of these sources
one may expect a CGB spectrum that displays significant changes in 
spectral index or even gaps.
The data (Figure 1) do not show a gap, but the spectral profile does
undergo a transition from the Seyfert regime to the blazar regime. The 
predicted gamma ray flux from SNIa fills the intermediate energy band at 
the approximately right level, but the detailed spectral shapes at the 
Seyfert-SNIa and SNIa-blazar interfaces are not fully explained by 
existing theoretical models. 

\section{Star Formation Rates}

The CGB flux from SNIa depends on the cosmic Star Formation Rate (SFR),
which is a sensitive function of redshift.
The first published SFR(z) for z = 0 to 5 \cite{Mada96}, now referred 
to as the ``Madau
Curve'', was based on the Hubble Deep Field - North (HDF-N). The data 
revealed a rapid
increase from z=0 to z $\sim$ 1 by a factor of 20 or so, then a 
decrease for higher z. SNIa
between z = 0 and z=2 contribute the bulk of the observed flux \cite{Wata99}.

\begin{figure}[htb]
\includegraphics[angle=90, scale=0.6]{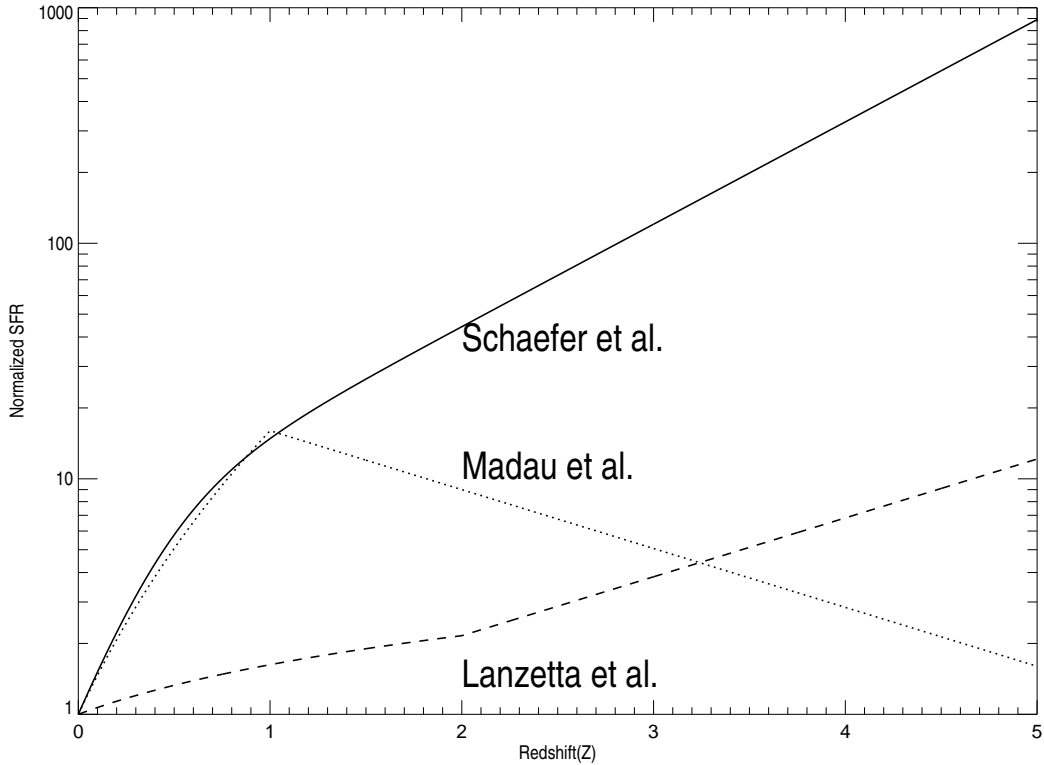}
\caption{Cosmic star formation rate per comoving volume normalized to
the present day rate density. The Madau-SFR(z) is contrasted
with rates from Lanzetta {\it et al.}\cite{Lanz02}, 
and Schaefer {\it et al.}\cite{Scha02}.The latter is derived from GRB observations.}
\label{fig:sfrs_plot}
\end{figure}

Several new estimates of SFR(z) have been published since the 
pioneering work of Madau.
Some studies suggest a flat SFR after z = 1, others argue for a
continued increase beyond z $\sim$ 1. 
A recent study \cite{Lanz02} using the Hubble Deep Field - 
South (HDF-S) suggests
that the original Madau estimate misses a dominant fraction of the 
ultraviolet luminosity
density of the universe at high redshift, 
and that the Madau curve ``at redshifts z $<$ 2 must be 
reduced by a factor of $\sim$ 2''.
Their new SFR(z) increases from z =0 to 2 by a factor of only 2, but 
more rapidly above z = 2
(Figure 2). We re-calculated the CGB using the Lanzetta-SFR, and find 
that the CGB flux is 
lower by a factor of 3 (at 1 MeV) than the flux derived from Madau's 
rate (Figure 3). Because other possible contributors to the CGB, as 
mentioned above, account for only a few $\%$ of the flux, we can not 
match the observed CGB with Lanzetta's SFR. Unless there is a class 
of not yet
recognized contributors to the MeV background, the Lanzetta-SFR appears 
inconsistent with the observed CGB. However, the SFR derived by 
Lanzetta et al. is only a lower limit. To make the predicted flux 
agree with the observed CGB the cosmic stellar production rate must be 
significantly larger than implied by Lanzetta et al..

\section{New SFR with CGB}

We also investigate the SFR obtained from Gamma-Ray Burst 
observations \cite{Scha02}, which increases dramatically with 
redshift beyond z $>$ 1 (in contrast to most SFR estimates, which 
either saturate or decline for z $>$ 1, see Figure 2). The CGB flux 
with the Schaefer-SFR not only matches the observed CGB flux in the 
MeV regime, but also reduces the apparent Seyferts-SNIa gap
in the few 100 keV regime (Figure 3).

\begin{figure}[htb]
\includegraphics[angle=90, scale=0.6]{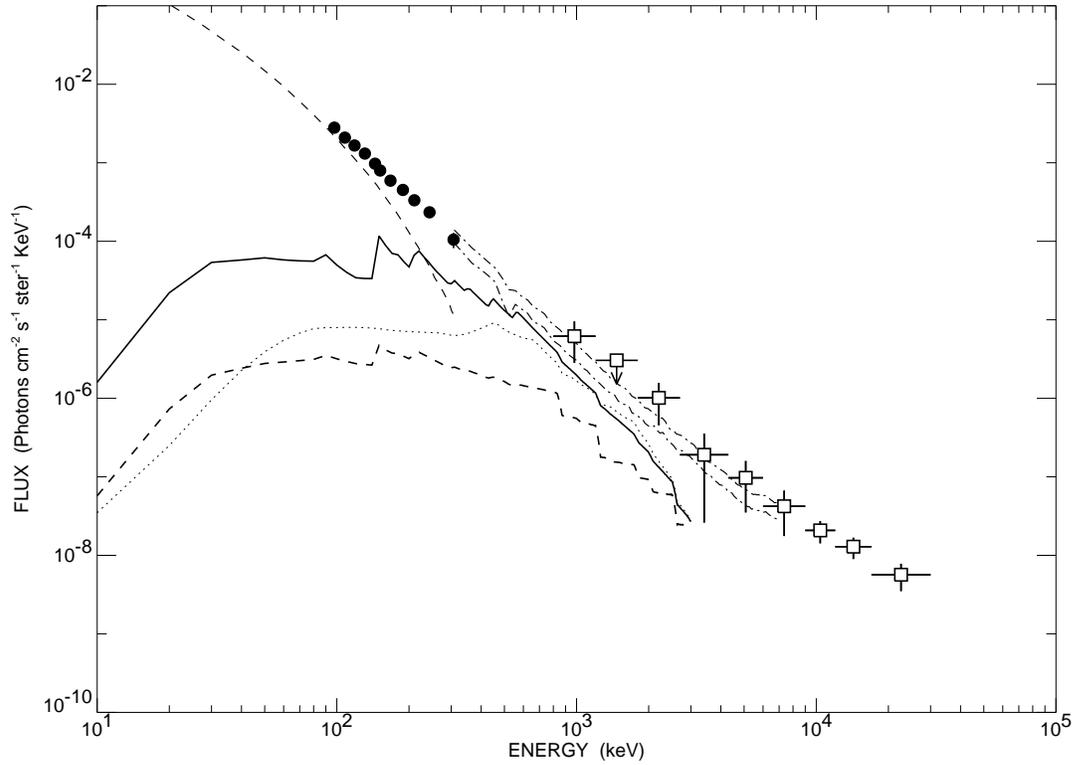}
\caption{The CGB obtained with the Schaefer-SFR(z) based on GRB 
observations (solid line). Dotted/dashed lines are 
the CGB obtained with the Madau-SFR and Lanzetta-SFR, respectively. 
Data are the same as shown in Figure 1.The Schaefer-SFR produces a better 
match to the observed CGB.}
\label{fig:cgb_schaefer}
\end{figure}

The number of gamma-ray bursts with measured redshifts is still too
small to consider the GRB-SFR to be the most reliable method of 
determining the cosmic star formation rate. However, the fact that the
CGB flux based on Schaefer's rate improves the match with the data 
does provide some additional support for the idea that GRBs trace 
the formation of massive stars (e.g.,Heger et al. \cite{Heger})
and thus provide a unique probe of the very high redshift universe
(Lamb $\&$ Reichart \cite{LR00}).

\section{Acknowledgement}
The authors thank B. Schaefer for providing his SFR(z) prior to 
publication. K.W. greatly appreciate for the generous support given 
by the NIC7 LOC.

\end{document}